\documentstyle[amssym, 11pt,aaspp4]{article}

\slugcomment{Accepted for publication in ApJ - 23rd March 1999}

\lefthead{Chadwick et al.}

\righthead{A Search for VHE Gamma Rays from Southern AGNs}

\begin{document}

\title{A Search for VHE Gamma Rays from AGNs Visible from the Southern
Hemisphere} 

\author{P.~M.~Chadwick, K.~Lyons, T.~J.~L.~McComb, K.~J.~Orford,
J.~L.~Osborne, S.~M.~Rayner, S.~E.~Shaw, and K.~E.~Turver}

\affil{Department of Physics, Rochester Building, Science Laboratories,
University of Durham, Durham, DH1 3LE, U.K.}

\authoremail{p.m.chadwick@dur.ac.uk}

\begin{abstract}

Observations have been made, using the University of Durham Mark 6 gamma
ray telescope, of the very high energy gamma ray emission from a number
of active galactic nuclei visible from the Southern hemisphere. Limits
are presented to the VHE gamma ray emission from 1ES
0323+022, PKS 0829+046, 1ES 1101--232, Cen A, PKS 1514--24, RXJ
10578--275, and 1ES 2316--423, both for steady long-term emission and
for outbursts of emission on timescales of 1 day.

\end{abstract}

\keywords{BL Lacertae objects: individual (PKS 1514--24, PKS 0829+046,
1ES 1011--232, 1ES 2316--423, 1ES 0323+022, RXJ 10578--275) galaxies:
individual (Cen A) --- galaxies: active --- gamma rays: observations}

\section{Introduction}

One of the most unexpected results in high energy astrophysics in the
last decade has been the discovery of high energy and very high energy
(VHE) emission from active galactic nuclei (AGNs). The EGRET detector on
board the {\it Compton Gamma Ray Observatory} established that BL Lacs
(predominantly radio selected) and flat-spectrum radio sources are
strong high energy gamma ray emitters, while X-ray selected BL Lacs have
been identified as a source of VHE gamma rays.

The first AGN to be identified as a VHE source was Mrk 421
(\cite{punch1992,macomb1995,petry1996}). Since then, Mrk 501
(\cite{quinn1996}, \cite{aharonian1997}), 1ES 2344+514
(\cite{catanese1998}) and PKS 2155--304 (\cite{chadwick1999}) have been
identified as VHE sources. Some of these objects have shown periods of
intense flaring activity on timescales as short as 15 minutes
(\cite{gaidos1996,aharonian1999}).

Previous surveys of VHE emission from AGNs
(\cite{kerrick1995,roberts1998,roberts1999,rowell1999}) have indicated
that only close X-ray selected BL Lacs are observable at TeV energies.
This is in accord both with theoretical models of gamma ray emission
from AGNs (for a recent review see \cite{ulrich1997}) and absorption of
VHE gamma rays on the cosmic infra-red background
(\cite{stecker1992,stecker1996,stecker1998}). Recent progress in the
understanding of blazars has shown that the distinction between
radio-selected (RBL) and X-ray selected (XBL) BL Lacs is not exact and
that there is a whole continuum between these two extremes
(\cite{ghisellini1998}). 

The Durham AGN dataset consists of observations of 10 AGNs made with the
Mark 6 telescope from 1996 to 1998. The discovery of VHE gamma rays from
PKS 2155--304 has already been reported; this is the most distant BL Lac
yet detected at these energies (\cite{chadwick1999}). Results from the
two close X-ray selected BL Lacs PKS 0548--322 and PKS 2005--489 will be
reported elsewhere (\cite{chadwick1999b}). Here we describe observations
of 1ES 0323+022, PKS 0829+046, RXJ 10578--275, 1ES 1101--232, Cen A, PKS
1514--24, and 1ES 2316--423, covering a range of classes of AGN. The
typical energy threshold for these observations is $ \sim 300~{\rm
to}~400 $ GeV. This is $\sim 5$ times lower than the typical threshold
of the CANGAROO telescope, which is $\sim 2$ TeV, which has also been
used to observe Southern hemisphere AGNs (\cite{roberts1999}).

\section{The Mark 6 Telescope}

The University of Durham Mark 6 atmospheric \v{C}erenkov telescope has
been in operation at Narrabri, NSW, Australia since July 1995. The
telescope is described in detail elsewhere (\cite{armstrong1997}). It
uses the imaging technique to separate VHE gamma rays from the cosmic
ray background, combined with a robust noise-free trigger based on the
signals from three parabolic flux detectors of 7 m diameter and aperture
f/1.0 mounted on a single alt-azimuth platform. A 109-element imaging
camera ($91 \times 0.25^\circ$ and $18 \times 0.5^{\circ}$ pixel size)
is mounted at the focus of the central flux detector, with low
resolution cameras each consisting of 19 pixels ($0.5^\circ$ pixel size)
mounted at the focus of the outer (left and right) flux collectors. The
telescope is triggered by demanding a simultaneous temporal (10 ns gate)
and spatial ($0.5^\circ$ aperture) coincidence of the \v{C}erenkov light
detected in the three cameras. This multiple-dish triggering system is
stable against variations in performance due to accidental coincidences,
and enables the telescope to detect low energy gamma rays with high
immunity from triggering by local muons. Initial work suggests that our
system has a detection probability for 100 GeV gamma rays of about 1\%
with the probability of detection rising slowly to about 40\% for 250
GeV gamma rays.

The Mark 6 telescope has been designed to provide stable operation which
allows the observation of weak DC sources. All detector packages are
thermally stabilised. The atmospheric clarity is continuously monitored
both using a far infra-red radiometer (\cite{buckley1999}) and an axial
optical CCD camera which allows the position and visual magnitude of
guide stars to be monitored. The gain and noise performance of the PMTs,
digitizer pedestals and associated electronics are continuously
monitored by:

\begin{enumerate}

\item triggering the telescope at random times using a nitrogen laser /
plastic scintillator / optical fibre light guide / opal diffuser system
to simulate \v{C}erenkov flashes and so enable flat-fielding, and

\item producing false triggers at random times to measure samples of the
background noise.

\end{enumerate}

\section{Observations}

Current VHE $\gamma$-ray observations of AGNs support the idea that it
is the XBLs which are the most promising sources of VHE emission, as
suggested by \cite{stecker1996}. The seven AGNs which are discussed in
this paper comprise three XBLs, two RBLs, one intermediate class object
and one close radio galaxy (Cen A) which has been detected previously as
a VHE $\gamma$-ray source. While RBLs are thought to be less promising
as VHE $\gamma$-ray sources than XBLs, observations in the VHE range
will help to confirm the fundamental differences between the XBLs and
RBLs. VHE $\gamma$-ray observations of BL Lacs have, in general,
concentrated on the closest objects, but we have sought to extend the
current redshift limit of $z = 0.117$ by observing more distant AGNs.
With an energy threshold of $\sim 300$ GeV, the Mark 6 Telescope is
well-suited to this task. In the case of one XBL, 1ES 1101--232, the VHE
$\gamma$-ray observations were made contemporaneously with {\it
BeppoSAX} observations.

Data were taken in 15-minute segments. Off-source observations were
taken by alternately observing regions of sky which differ by $\pm~15$
minutes in right ascension from the position of the object to ensure
that the on and off segments possess identical azimuth and zenith
profiles. This off-source -- on-source -- on-source -- off-source
observing pattern is routinely used to eliminate any first order changes
in count rate due to any residual secular changes in atmospheric
clarity, temperature etc.

Data were accepted for analysis only if:

\begin{enumerate}

\item the sky was clear and stable, and

\item the gross counting rates in each on-off pair were consistent at
the $2.5~\sigma$ level.

\end{enumerate}

A total of 54 hours of on-source observations under clear skies of 7
objects was completed, and an observing log is shown in Table
\ref{observing_log}.

\section{Data Analysis}

Routine reduction and analysis of accepted data comprises the following
steps:

\begin{enumerate}

\item calibration of the gains and pedestals of all 147 PMTs with their
associated digitizer electronics within each 15 minute segment, using
the embedded laser and false coincidence events,

\item software padding of the data to equalize the effects of on- and
off-source sky noise on data selection (\cite{cawley1993};
\cite{fegan1997}),

\item identification of the location of the source in the
camera's field of view for each event, using the axial CCD camera,

\item a calculation of the spatial moments of each shower image relative
to the source position, and

\item rejection of events containing an image which would be unlikely to
be produced by gamma rays.

\end{enumerate}

Events considered suitable for parameterization are those which are
confined within the sensitive area of the camera (i.e. within
$1.1^{\circ}$ of the centre of the camera) and which contain sufficient
information for reliable image analysis, i.e. which have {\it SIZE} $ >
500$ digital counts, where there are $\sim 3$ digital counts per 1
photoelectron, and typically 200 digital counts are produced by a 125
GeV gamma ray. 

The \v{C}erenkov images recorded with the Mark 6 telescope are
parameterized after \cite{hillas1985}. These image parameters allow
discrimination between the elliptical images, the major axes of which
point towards the source direction, produced by a $\gamma$-ray shower
and the broader, more irregular images produced by a hadronic shower. In
addition, a measure of the difference between the positions of the
centroids of the light recorded by the left and right flux collectors of
the Mark 6 telescope provides a further discriminant, $D_{\rm dist}$
(\cite{chadwick1998a}). Gamma rays are identified by selecting events on
the basis of image shape and left/right fluctuation, and then plotting
the number of events as a function of the pointing parameter {\it
ALPHA}; $\gamma$-ray events from a point source will appear as an excess
of events at small values of {\it ALPHA}. In this case, we define the
$\gamma$-ray domain as {\em ALPHA} $< 22.5^{\circ}$.

The selection criteria applied to these data are summarized in Table
\ref{select_table}. They constitute a standard set of criteria developed
from our successful observations of PKS 2155--304, and include allowance
for the variation of image parameters with event size. They are routinely
applied to data from all objects recorded at zenith angles less than
$45^{\circ}$, which is the case for all the observations reported here.

\section{Results}

\begin{figure}[tbh]

\epsscale{1.0}
\plotone{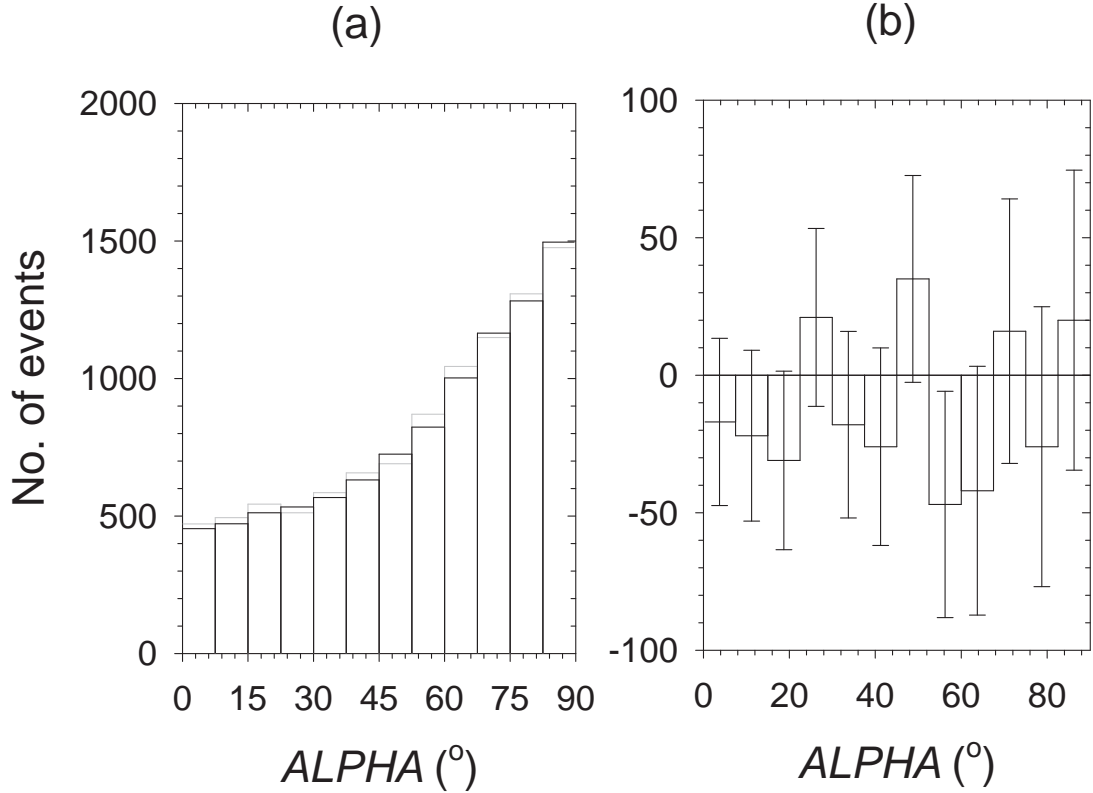}

\caption{(a) The {\em ALPHA} distributions ON and OFF source for 1ES
2316--423. The dotted line refers to the OFF source data. (b) The
difference in the {\it ALPHA} distributions for ON and OFF source
events. \label{alpha_plot} }

\end{figure}

The dataset for each source has been tested for the presence of gamma
ray signals. Typical results of the application of the cuts described
above to one object (1ES 2316--423) are shown in Table
\ref{events_table}, and the corresponding {\em ALPHA}-plot is shown in
Figure \ref{alpha_plot}. The flux limits from the seven AGNs are
summarised in Table \ref{results_table}. They are all $3\sigma$ flux
limits, based on the maximum likelihood ratio test
(\cite{gibson1982,li1983}). The threshold energy for the observations
has been estimated on the basis of preliminary simulations, and is in
the range 300 to 400 GeV for these objects, depending on the object's
elevation. The collecting areas which have been assumed, again from
simulations, are $ 5.5 \times 10^{8} ~{\rm cm}^{2}$ at an energy
threshold of 300 GeV and $ 1.0 \times 10^{9} ~{\rm cm}^{2}$ at an energy
threshold of 400 GeV. These are subject to systematic errors estimated
to be $\sim 50 \%$. We have assumed that our current selection
procedures retain $\sim 20 \%$ of the $\gamma$-ray signal, which is
subject to a systematic error of $\sim 60$ \%.



We have also searched our dataset for $\gamma$-ray emission on
timescales of $\sim 1$ day. The search for enhanced emission has been
conducted by calculating the on-source excess after the application of
our selection criteria for the pairs of on/off observations recorded
during an individual night. A typical observation comprising 6 on/off
pairs of observations (1.5 hours of on-source observations) yields a
flux limit of $\sim 1 \times 10^{-10} {\rm ~cm}^{-2} {\rm ~s}^{-1}$ at
300 GeV. Conversely, had any of the objects on which we report here
produced a 15-minute flare similar to that seen from Mrk 421 with the
Whipple telescope on 1996 May 7 (\cite{gaidos1996}), it would be
detected with the Mark 6 telescope at a significance of around
$7~\sigma$. There is no evidence for any flaring activity. 

\subsection{Cen A}

Centaurus A (NGC 5128) is the closest radio-loud active galaxy to Earth,
at a distance of 5 Mpc ($z = 0.008$), and is often described as the
prototype Fanaroff-Riley Class I low luminosity radio galaxy. It was
identified as a TeV source in the early days of VHE gamma ray astronomy
(\cite{grindlay1975}), with a flux of $4.4 \pm 1.0 \times 10^{-11} {\rm
~cm}^{-2} {\rm ~s}^{-1}$ at an energy threshold of 300 GeV when the
object was in an X-ray high state. Observations of Cen A were also made
with the University of Durham Mark 3 telescope which placed a $3 \sigma$
flux limit of $7.8 \times 10^{-11} {\rm ~cm}^{-2} {\rm ~s}^{-1}$ at a
similar energy threshold (\cite{carraminana1990}). The X-ray state of
Cen A at the time of these observations was unknown. Observations of Cen
A made in 1995 March/April with the CANGAROO telescope have resulted in
a $3 \sigma$ flux limit of $4.66 \times 10^{-12} {\rm ~cm}^{-2}{\rm
~s}^{-1}$ at $E > 1.5$ TeV for an extended source centred on Cen A
(\cite{rowell1999}). There has also been a report at UHE energies of an
excess in cosmic ray showers from the direction of Cen A
(\cite{clay1984}). EGRET observations have recently been used to
identify Cen A as a source of GeV gamma rays (\cite{sreekumar1999}),
thus providing the first evidence for emission in the 30 -- 10000 MeV
energy range from a source with a confirmed large-inclination jet. 

The observations of Cen A made with the Mark 6 imaging telescope
reported here provide a flux limit of $5.2 \times 10^{-11} {\rm
~cm}^{-2} {\rm ~s}^{-1}$. {\em BeppoSAX} observations made in 1997
February, approximately two weeks before the commencement of our
observations, show the source to have been in a low state, lower by at
least a factor of $\sim 5$ than the outburst in 1974 -- 5
(\cite{grandi1999}). {\em RXTE} observations taken contemporaneously
with our data confirm that Cen A was in a low state in 1997
March.\footnote{Available on the web at
\mbox{http://space.mit.edu/XTE/asmlc/cena.html}.} If, as seems to be the
case in other AGNs, the X-ray and VHE $\gamma$-ray emission from Cen A
are correlated, then it may not be surprising that we detected no VHE
emission in 1997 March.

\subsection{PKS 0829+046}

PKS 0829+046, also known as OJ049, has a redshift of 0.18
(\cite{falomo1991}). It was detected with both {\em HEAO-1} as an X-ray
source (\cite{dellaceca1990}) and the EGRET instrument as a GeV
$\gamma$-ray source
(\cite{fichtel1994,vonmontigny1995,mattox1997,mukherjee1997}). It has a
large radio flux and is therefore classified as an RBL
(\cite{ciliegi1993}); this suggests it is unlikely to be a detectable
VHE $\gamma$-ray source. The present VHE limit is $4.7 \times 10^{-11}
{\rm ~cm}^{-2} {\rm ~s}^{-1}$ for $E > 400$ GeV.

\subsection{PKS 1514--24}

Misidentified initially as AP Librae, PKS 1514--24 was one of the first
radio-detected BL Lacs (\cite{bolton1965}). It has a redshift of 0.049,
and although detected by {\em EXOSAT} (\cite{schwartz1983}), its
relatively small X-ray flux classifies it as an RBL
(\cite{ciliegi1993}). Phase 1 observations with the EGRET detector on
board {\em CGRO} resulted in an upper limit for the object of $7 \times
10^{-8} {\rm ~cm}^{-2} {\rm ~s}^{-1}$ at $E > 100$ MeV
(\cite{fichtel1994}) nor does it appear in the 3rd EGRET catalog
(\cite{hartman1999}). The VHE limit presented here is $3.7 \times
10^{-11} {\rm ~cm}^{-2}{\rm ~s}^{-1}$ for $E > 300$ GeV.

\subsection{1ES 2316--423}

1ES 2316--423 ($z = 0.055$) was originally classified as a radio
selected BL Lac (known as PKS 2316--423, see e.g. \cite{stickel1991}),
but recently \cite{perlman1998} have identified this object as an
intermediate case whose high energy emission could be expected to
extend up to VHE energies. The CANGAROO telescope has observed this
object but detected no VHE emission, placing a $2 \sigma$ upper limit of
$1.2 \times 10^{-12} {\rm~cm}^{-2} {\rm~s}^{-1}$ at a threshold energy
of $\sim 2$ TeV in July 1996 (\cite{roberts1998}). The present
measurement indicates a flux limit at the $3 \sigma$ level of $4.5
\times 10^{-11} {\rm ~cm}^{-2} {\rm ~s}^{-1}$ at $E > 300$ GeV. Assuming
an integral spectral index of $\sim 1.5$ (c.f. Mrk 421 and Mrk 501), this
corresponds to a $3\sigma$ flux limit at $E > 2$ TeV of $\sim 2.6 \times
10^{-12} {\rm ~cm}^{-2} {\rm ~s}^{-1}$, comparable with the $2~\sigma$
flux limit from the CANGAROO experiment.

\subsection{1ES 1101--232}

1ES 1101--232 is an XBL with a redshift of 0.186. It has been detected
using both the {\em HEAO-1} and {\em Einstein} satellites
(\cite{dellaceca1990,perlman1996}). Phase one EGRET observations
resulted in an upper limit of $6 \times 10^{-8} {\rm ~cm}^{-2} {\rm
~s}^{-1}$ at $E > 100$ MeV (\cite{fichtel1994}). It was detected with
the {\it BeppoSAX} satellite in 1997 (\cite{wolter1998}), and our
observations of this XBL were made contemporaneously with a {\it
BeppoSAX} campaign on the object. Indications are that the X-ray flux
from 1ES 1101--232 was $\sim 30\%$ lower during our observations than in
1997 (Wolter, private communication). Our flux limit is $3.7 \times
10^{-11} {\rm~cm}^{-2}{\rm~s}^{-1}$ at $E > 300$ GeV.

\subsection{RXJ 10578--275}

The {\em Rosat} source RXJ 10578--275 was initially identified as a
potential BL Lac from its optical characteristics (\cite{bade1994}). It
has a redshift of 0.092 and is classified as an XBL. Our flux limit is
$8.2 \times 10^{-11} {\rm ~cm}^{-2}{\rm ~s}^{-1}$ at $E > 300$ GeV.

\subsection{1ES 0323+022}

1ES 0323+022 has been detected using both the {\em HEAO-1} and {\em
Einstein} satellites (\cite{dellaceca1990}, \cite{perlman1996}). It is
an XBL with a redshift of 0.147, and has a spectrum which is very
similar to the archetypal XBL PKS 2155--304 (\cite{giommi1995}). EGRET
phase one observations resulted in an upper limit of $ 6 \times 10^{-8}
{\rm ~cm}^{-2} {\rm ~s}^{-1}$ (\cite{fichtel1994}). \cite{stecker1996}
predict a flux at $E > 300$ GeV of $4.0 \times 10 ^{-12}{\rm
~cm}^{-2}{\rm ~s}^{-1}$. Our flux limit of~$3.7 \times 10^{-11}
{\rm~cm}^{-2}{\rm~s}^{-1}$ at $E > 300$ GeV is considerably higher than
this prediction and so it is not in conflict.

\section{Discussion}

Whilst the interpretation of VHE upper limits from BL Lacs is
complicated by the lack of a complete theory of VHE $\gamma$-ray
emission from AGNs, \cite{stecker1996} have predicted the TeV fluxes
from a range of objects, one of which (1ES 0323+022), is included in the
present work. The expected fluxes from the other XBLs included in this
paper may be estimated on the basis of the work of \cite{stecker1996}
and \cite{stecker1998b} using the simple relation $\nu_{x}F_{x} \sim
\nu_{\gamma}F_{\gamma}$ and the published X-ray fluxes. We estimate that
the 300 GeV fluxes of 1ES 1101--232, 1ES 2316--423 and RXJ 10578--275
would be $2.0 \times 10 ^{-11}{\rm ~cm}^{-2}{\rm ~s}^{-1}$, $1.5 \times
10 ^{-12}{\rm ~cm}^{-2}{\rm ~s}^{-1}$, and $3.3 \times 10 ^{-12}{\rm
~cm}^{-2}{\rm ~s}^{-1}$ respectively, taking into account photon-photon
absorption using the recent determination of $\gamma$-ray opacity by
\cite{stecker1998b}. All these suggested fluxes are lower than the flux
limits reported here. However, the lack of contemporaneous X-ray
measurements in the case of most of our observations limits the
usefulness of these predictions and emphasises the importance of
simultaneous X-ray and $\gamma$-ray observations and multiwavelength
campaigns. In the case of the RBLs, an extended observation of PKS
1514--24, a close RBL, lends support to the suggestion that RBLs are not
strong VHE $\gamma$-ray emitters. 

Our observations of Cen A were made when it was in an X-ray low state,
in contrast to the earlier VHE detection of Cen A reported by
\cite{grindlay1975}, which was made when Cen A was in X-ray outburst.
Further VHE $\gamma$-ray observations during an X-ray high state would
be desirable.

\section{Conclusions}

The Durham University Mark 6 Telescope has been used to make
observations of 7 close AGNs: 1ES 0323+022 (XBL, $z = 0.147$), PKS
0829+046 (RBL, $z = 0.18$), RXJ 10578--275 (XBL, $z = 0.092$) 1ES
1101--232 (XBL, $z = 0.186$), Cen A (low luminosity radio galaxy, $z =
0.0089$), PKS 1514--24 (RBL, $z = 0.049$) and 1ES 2316--423
(transitional BL Lac, $z = 0.055$). We find no evidence for either
steady or flaring emission of $\gamma$-rays above 300 -- 400 GeV. The
flux limits are in excess of the fluxes predicted on the basis of the
simple model of \cite{stecker1996}. The flux limit derived for 1ES
0323+022 ($3.7 \times 10^{-11}{\rm~cm}^{-2}{\rm~s}^{-1}$) is not in
conflict with the specific prediction of \cite{stecker1996} ($4.0 \times
10^{-12}{\rm~cm}^{-2}{\rm~s}^{-1}$).

\acknowledgements

We are grateful to the UK Particle Physics and Astronomy Research
Council for support of the project and the University of Sydney for the
lease of the Narrabri site. The Mark 6 telescope was designed and
constructed with the assistance of the staff of the Physics Department,
University of Durham. The efforts of Mrs. S. E. Hilton and Mr. K.
Tindale are acknowledged with gratitude. We would like to thank Anna
Wolter for providing us with information about {\it BeppoSAX}
observations of 1ES 1011--232 in advance of publication. This paper uses
quick look results provided by the ASM/{\it RXTE} team and uses the
NASA/IPAC Extragalactic database (NED), which is operated by the Jet
Propulsion Laboratory, Caltech, under contract with the National
Aeronautics and Space Administration.


\newpage

\begin{table*}
\begin{center}
\begin{tabular}{@{}lcc}
\tableline
\tableline
Object & Date & No. of \\
& & ON source scans \\
\tableline

Cen A & 1997 March 08 & 5 \\
Cen A & 1997 March 10 & 5 \\
Cen A & 1997 March 11 & 6 \\
Cen A & 1997 March 12 & 6 \\
Cen A & 1997 March 13 & 5 \\
PKS 0829+046 & 1996 March 15 & 3 \\
PKS 0829+046 & 1996 March 17 & 6 \\
PKS 0829+046 & 1996 March 18 & 7 \\
PKS 1514--24 & 1996 April 14 & 2 \\
PKS 1514--24 & 1996 April 15 & 6 \\
PKS 1514--24 & 1996 April 17 & 6 \\
PKS 1514--24 & 1996 April 18 & 7 \\
PKS 1514--24 & 1996 April 19 & 10 \\
PKS 1514--24 & 1996 April 20 & 6 \\
PKS 1514--24 & 1996 April 21 & 6 \\
PKS 1514--24 & 1996 April 22 & 8 \\
1ES 2316--423 & 1997 August 26 & 2 \\
1ES 2316--423 & 1997 August 27 & 2 \\
1ES 2316--423 & 1997 August 29 & 8 \\
1ES 2316--423 & 1997 August 30 & 13 \\
1ES 2316--423 & 1997 August 03 & 11 \\
1ES 2316--423 & 1997 September 06 & 4 \\
1ES 1101--232 & 1998 May 19 & 6 \\
1ES 1101--232 & 1998 May 20 & 3 \\
1ES 1101--232 & 1998 May 21 & 7 \\
1ES 1101--232 & 1998 May 22 & 6 \\
1ES 1101--232 & 1998 May 23 & 4 \\
1ES 1101--232 & 1998 May 24 & 4 \\
1ES 1101--232 & 1998 May 25 & 8 \\
1ES 1101--232 & 1998 May 26 & 8 \\
1ES 1101--232 & 1998 May 27 & 6 \\
RXJ 10578--2753 & 1996 March 20 & 7 \\
RXJ 10578--2753 & 1996 March 21 & 6 \\
RXJ 10578--2753 & 1996 March 22 & 2 \\
1ES 0323+022 & 1996 September 14 & 4 \\
1ES 0323+022 & 1996 September 15 & 4 \\
1ES 0323+022 & 1996 September 17 & 7 \\

\tableline
\end{tabular}
\end{center}

\caption{Observing log for observations of active galactic nuclei made
with the University of Durham Mark 6 Telescope.}

\label{observing_log} 
\end{table*}


\begin{table*}

\begin{center}

\begin{tabular}{@{}lccccc}
\tableline
\tableline
Parameter&Ranges&Ranges&Ranges&Ranges&Ranges\\
\tableline

{\it SIZE}
(d.c.)&$500-800$&$800-1200$&$1200-1500$&$1500-2000$&$2000-10000$\\
{\it
DISTANCE}&$0.35^{\circ}-0.85^{\circ}$&$0.35^{\circ}-0.85^{\circ}$&$0.
35^{\circ}-0.85^{\circ}$&$0.35^{\circ}-0.85^{\circ}$&$0.35^{\circ}-0.
85^{\circ}$\\
{\it
ECCENTRICITY}&$0.35-0.85$&$0.35-0.85$&$0.35-0.85$&$0.35-0.85$&$0.35-0.
85$\\
{\it WIDTH}&$ < 0.10^{\circ}$&$ < 0.14^{\circ}$&$ < 0.19^{\circ}$&$ <
0.32^{\circ}$&$ < 0.32^{\circ}$\\
{\it CONCENTRATION}&$ < 0.80$&$ < 0.70$&$< 0.70$&$ < 0.35$&$< 0.25$\\
$D_{\rm dist}$&$ < 0.18^{\circ}$&$ < 0.18^{\circ}$&$ < 0.12^{\circ}$&$ <
0.12^{\circ}$&$ < 0.10^{\circ}$\\
\tableline

\end{tabular}

\end{center}

\caption{The image parameter selections applied to the data.}

\label{select_table}

\end{table*}


\begin{table*}
\begin{center}
\begin{tabular}{@{}lrrrr}
\tableline
\tableline
& ON & OFF & Difference & Significance\\
\tableline

Number of events&218541&220531&-1990&-3.0$\sigma$\\
Number of size and distance selected events&121334&121426&-92&-0.19$\sigma$\\
Number of shape selected events&5454&5542&-88&-0.85$\sigma$\\
Number of shape and $ALPHA$ selected events&1438&1508&-70&-1.29$\sigma$\\
\tableline

\end{tabular}

\end{center}

\caption{The result of the application of image parameter selections to
the data from 1ES 2316--423.}

\label{events_table}

\end{table*}

\begin{table*}
\begin{center}
\begin{tabular}{@{}lcc}
\tableline
\tableline
Object & Estimated & Flux Limit \\
& Threshold (GeV) & $ (\times 10^{-11} {\rm~cm}^{-2} {\rm~s}^{-1})$\\
\tableline

Cen A & 300 & 5.2 \\
PKS 0829+046 & 400 & 4.7 \\
PKS 1514--24 & 300 & 3.7 \\
1ES 2316--423 & 300 & 4.5 \\
1ES 1101--232 & 300 & 3.7 \\
RXJ 10578--275 & 300 & 8.2 \\
1ES 0323+022 & 400 & 3.7 \\

\tableline
\end{tabular}
\end{center}

\caption{Flux limits for observations of active galactic nuclei made
with the University of Durham Mark 6 Telescope.}

\label{results_table} 
\end{table*}


\begin{thebibliography}{}


\bibitem[Aharonian et al. 1997]{aharonian1997} Aharonian, F. A. et al.
1997 A\&A, 327, L5

\bibitem[Aharonian et al. 1999]{aharonian1999} Aharonian, F. A., et al.
1999 A\&A, 342, 69

\bibitem[Armstrong et al. 1999]{armstrong1997} Armstrong, P., et al.
1999, Exp. Astron., in press

\bibitem[Bade, Fink, \& Engels 1994]{bade1994} Bade, N., Fink, H. H., \&
Engels, D. 1994, A\&A, 286, 388. 

\bibitem[Bolton, Clarke, \& Ekers 1965]{bolton1965} Bolton, J. G.,
Clarke, M. G., \& Ekers, R. D. 1965, Aust. J. Phys., 18, 627 

\bibitem[Buckley et al. 1999]{buckley1999} Buckley, D. J., Dorrington,
M. C., Edwards, P. J., McComb, T. J. L., Tummey, S. P., \& Turver, K. E.
1999, Exp. Astron., submitted

\bibitem[Carraminana et al. 1990]{carraminana1990} Carraminana, A., et
al. 1990, A\&A, 228, 327

\bibitem[Catanese et al. 1998]{catanese1998} Catanese, M. A., et al.
1998, \apj, 501, 616

\bibitem[Cawley 1993]{cawley1993} Cawley, M. F. 1993,
Towards a Major Atmospheric Cerenkov Detector II ed. R. C. Lamb, 176

\bibitem[Chadwick et al. 1998]{chadwick1998a} Chadwick, P. M., et al.
1998, \apj, 503, 391

\bibitem[Chadwick et al. 1999a]{chadwick1999} Chadwick, P. M., et al.
1999a, \apj, in press

\bibitem[Chadwick et al. 1999b]{chadwick1999b} Chadwick, P. M., et al.
1999b, in preparation

\bibitem[Ciliegi, Bassani, \& Caroli 1993]{ciliegi1993} Ciliegi, P.,
Bassani, L., \& Caroli, E. 1993 \apjs, 85, 111

\bibitem[Clay, Gerhardy, \& Liebing 1984]{clay1984} Clay, R. W.,
Gerhardy, P. R., \& Liebing, D. F. 1984, Australian J. Phys., 37, 91

\bibitem[Della Ceca et al. 1990]{dellaceca1990} Della Ceca, R., Palumbo,
G. G. C., Persic, M., Boldt, E. A., Marshall, E. E. \& de Zotti, G.
1990, \apjs, 72, 471

\bibitem[Falomo 1991]{falomo1991} Falomo, R. 1991, \aj, 102, 1991

\bibitem[Fegan 1997]{fegan1997} Fegan, D. J. 1997, J. Phys. G. Nucl.
Part. Phys., 23, 1013

\bibitem[Fichtel et al. 1994]{fichtel1994} Fichtel, C. E., et al. 1994,
\apjs, 94, 551 

\bibitem[Gaidos et al. 1996]{gaidos1996} Gaidos, J. A., et al. 1996,
\nat, 383, 319

\bibitem[Ghisellini et al. 1998]{ghisellini1998} Ghisellini, G., et al.
1998, \mnras, 301, 451

\bibitem[Gibson et al. 1982]{gibson1982} Gibson, A. I., et al. 1982,
Proc. Intl. Workshop on Very High Energy Gamma Ray Astro., Bombay: Tata
Institute, ed. P. V. Ramana Murthy \& T. C. Weekes, 97

\bibitem[Giommi, Ansari, \& Nicol 1995]{giommi1995} Giommi, P., Ansari,
S. G., \& Micol, A. 1995, A\&AS, 109, 267

\bibitem[Grandi, Urry, \& Maraschi 1999]{grandi1999} Grandi, P., Urry,
C.M., \& Maraschi, L. 1999, astro-ph/9901266.

\bibitem[Grindlay et al. 1975]{grindlay1975} Grindlay, J. E., Helmken,
H. F., Hanbury Brown, R., Davis, J., \& Allen, L. R. 1975, \apjl, 197,
L9

\bibitem[Hartman et al. 1999]{hartman1999} Hartman, R. C., et al. 1999,
\apjs, in press

\bibitem[Hillas 1985]{hillas1985} Hillas, A. M. 1985, Proc. 19th Int.
Cosmic Ray Conf., (La Jolla), 3, 445

\bibitem[Kerrick et al. 1995]{kerrick1995} Kerrick, A. D., et al. 1995,
\apj, 452, 588

\bibitem[Li \& Ma 1983]{li1983} Li, T. P., \& Ma, Y. Q. 1983, \apj, 272,
317

\bibitem[Macomb et al. 1995]{macomb1995} Macomb, D. J. et al. 1995,
\apjl, 449, L99

\bibitem[Mattox et al. 1997]{mattox1997} Mattox, J. R., Schachter, J.,
Molnar, L., Hartman, R. C., \& Patnaik, A. R. 1997, \apj, 481, 95

\bibitem[Mukherjee et al. 1997]{mukherjee1997} Mukherjee, R. et al.
1997, \apj, 490, 116

\bibitem[Perlman et al. 1996]{perlman1996} Perlman, E. S. et al. 1996,
\apjs, 104, 251

\bibitem[Perlman et al. 1998]{perlman1998} Perlman, E. S. et al. 1998,
\aj, 115, 1253

\bibitem[Petry et al. 1996]{petry1996} Petry, D., et al. 1996, A\&A, 311,
L13

\bibitem[Punch et al. 1992]{punch1992} Punch, M., et al. 1992, \nat,
358, 477

\bibitem[Quinn et al. 1996]{quinn1996} Quinn, J., et al. 1996, \apjl,
456, L83

\bibitem[Roberts et al. 1998]{roberts1998} Roberts, M. D., et al. 1998,
A\&A, 337, 25

\bibitem[Roberts et al. 1999]{roberts1999} Roberts, M. D., et al. 1999,
A\&A, 343, 691

\bibitem[Rowell et al. 1999]{rowell1999} Rowell, G. P., et al., 1999,
astro-ph/9901316

\bibitem[Schwartz \& Ku 1983]{schwartz1983} Schwartz, D. A. \& Ku, W. H.
1983, \apj, 266, 459

\bibitem[Sreekumar et al. 1999]{sreekumar1999} Sreekumar, P., Bertsch,
D. L., Hartman, R. C., Nolan, P. L., \& Thompson, D. J. 1999,
astro-ph/9901277

\bibitem[Stecker 1998]{stecker1998b} Stecker, F. W. 1998,
astro-ph/9812286

\bibitem[Stecker \& de Jager 1998]{stecker1998} Stecker, F. W., \& de
Jager, O. C. 1998, A\&A, 334, L85

\bibitem[Stecker, de Jager, \& Salamon 1992]{stecker1992} Stecker, F.
W., de Jager, O. C., \& Salamon, M. H. 1992, \apjl, 390, L49

\bibitem[Stecker, de Jager, \& Salamon 1996]{stecker1996} Stecker, F.
W., de Jager, O. C., \& Salamon, M. H. 1996, \apjl, 473, L75

\bibitem[Stickel et al. 1991]{stickel1991} Stickel, M., Padovani, P.,
Urry, C. M., Fried, J. W., \& K\"{u}hr, H. 1991, \apj, 374, 431

\bibitem[Ulrich, Maraschi, \& Urry 1997]{ulrich1997} Ulrich, M.-H.,
Maraschi, L., \& Urry, C. M. 1997, Annu. Rev. Astron. Astrophys., 35,
445

\bibitem[von Montigny et al. 1995]{vonmontigny1995} von Montigny, C. et
al. 1995, \apj, 440, 525

\bibitem[Wolter et al. 1998]{wolter1998} Wolter, A. et al. 1998, A\&A,
335, 899


\end{thebibliography}
\end{document}